# Open Source Energy Simulation for Elementary School

Sze Yee Lye [a*], Loo Kang Wee[a] & Yong Cheng Ong[a]
[a]*Educational Technolgy Division, Ministry of Education,  Singapore*
*lye_sze_yee@moe.gov.sg

**Abstract:** With the interactivity and multiple representation features, computer simulations lend itself to the guided inquiry learning. However, these simulations are usually designed for post-elementary students. Thus, the aim of this study is to investigate how the use of guided inquiry approach with customized energy simulation can improve students' understanding of this topic. In this ongoing research, the case study is adopted. In the first phase of the study, we have modified open source energy simulation based on principles for reducing extraneous processing, existing energy simulation and guided inquiry approach. The modified simulation is sent to teachers for evaluation and the feedback is encouraging. In the next phase of the study, the guided inquiry lesson package involving the energy simulation would be designed and deployed in an elementary classroom. Multiple data sources would be collected to seek a deeper understanding on how this learning package can possibly impact students' understanding of the physics concepts.

**Keywords:** simulation, elementary school, physics, science, open source,

**Introduction**

With the proliferation of computing devices in the classrooms, science teachers are increasingly using technology (e.g., data loggers and computer simulations) to create meaningful learning experiences [1] for the students. Computer simulations, in particular, have gained popularity among the science educators as there is a wealth of easily available of online free and realistic computer [2]. Some examples of computer simulations include PhET's interactive Science Simulations at http://phet.colorado.edu/ , Open Source Physics Simulations at http://www.opensourcephysics.org/ and NTNUJAVA Virtual Physics Laboratory at http://www.phy.ntnu.edu.tw/ntnujava/.  A computer simulation is "a program that contains a model of a system (natural or artificial; e.g., equipment) or a process" (p. 180)[3]. Such simulation can also accept inputs from the users and present the computational results in multiple representations like graphs or tables [4]. By providing guidance in inquiry-based activities, computer simulations can be adopted in guided inquiry approach.[5]

Despite the potential of the simulations for use in guided inquiry learning, there  is paucity of such research conducted in elementary school . Searches done in  the three databases, *Academic Search Premier*, *PsycARTICLES* and *PsycINFO*, using descriptors "science simulations" and "elementary school" only yields 2 results [6,7]. This might  be due to the lack of  online science simulations created especially for elementary school. Moreover, little research has been done to investigate how other factors  (i.e., teachers' facilitation and classroom settings) may impact the students' learning using such simulations[2]. To address this research gap, this paper examines the use of customized energy simulation in elementary classroom. The case study approach is adopted and is guided by the following questions :  (1) What are the inquiry learning principles (e.g., level of teacher facilitation)  that can improve students' understanding of energy concepts? (2)

What are the design features of the computer simulations that help the students to understand energy concepts? In this study, the energy simulation is modified based on one of the many physics simulations found in Open Source Physics Simulations[8]. This is made possible as open source applications are characterized by the access to the source code and free distribution of the application[9]. This simulation is customized in such a way that it can : (1) provide more support for the inquiry-based activities and (2) reduce cognitive load. Such freeing up of cognitive resources is likely to enhance the guided learning experience [5].

## 1. Computer Simulations in Guided Inquiry

Inquiry based approach has always been popular in science education [10]. Learners are situated in an inductive learning mode, in which students are the "active agent in the process of knowledge acquisition" [2]. Usually, guided inquiry is preferred as the absence or little scaffolding may hinder the students' learning [2]. During such learning, the students are "posing and responding to questions, designing investigations, and evaluating and communicating their learning". [11]

Guided inquiry can be implemented with computer simulations supporting inquiry-based activities. Firstly, as cognitive tool, simulations will help processing of the data (e.g., the representation of data in table or graphic form)[12]. With this ability to share cognitive load, the simulations enable them to focus on higher order thinking skills (e.g., evaluating findings and designing investigation). Moreover, the simulation offers multiple representations (e.g., word, pictures, diagrams, graphs and table of values) of the same or related concepts which help the learners in responding to questions, evaluating and communicating their conclusion[13]. Such representations can foster deeper understanding of the science concepts as the learners can "integrate information from the various representations to achieve insights that otherwise would be difficult to achieve with only a single representation" [14]. Lastly, computer simulations offer interactivity, in which experimental variables can be manipulated [15]. Such affordance allows the learner to design investigation and evaluate their conclusion.

## 2. Open Source Energy Physics simulation

The topic on energy (e.g. kinetic energy and potential energy) is an important and difficult topic for the students [16]. The use of guided inquiry with energy simulation has the potential to solve this problem. Hence, we choose to modify the existing open source roller coaster simulation [8] so that elementary school students can investigate the energy concepts in a virtual setting. The original roller coaster simulation is designed with equations that model closely the energy concept. This makes the simulation fairly realistic and is not oversimplified such that students will have misconceptions [17]. There are, of course, other good energy-related simulations on the web . As such simulations are designed with older students in mind, the simulations may contain extra information which makes them unsuitable for use in elementary school.

In this study, we have successfully customized an open source roller coaster simulation using Easy Java Simulation (EJS). EJS at http://fem.um.es/Ejs/, free authoring toolkit for creating physics simulation, is part of the Open Source Physics project which aims to spread the use of open source code for physics simulations. Continuing the spirit of open source, this remixed simulation is shared online so that others can further refine or benefit from it [18] . Such spirit resonate with us, the science educators, as both open source applications and scientific knowledge are "built on the concept of creating shared knowledge and the desire to have one's work adopted by the scientific or computer-using community" [19].

## 3. Customization the Open Source Energy Simulation

As this simulation is meant for students in middle and high school, there is a need to customize the roller coaster simulation [8]. In the customization process, we are guided by: (1) the principles for reducing extraneous processing, (2) the existing good energy simulations and (3) inquiry learning principles. We used some of the principles(See Table 1) for reducing such extraneous processing outlined by Mayer, who had made significant contribution in multimedia learning [20] . Such reduction in cognitive load is vital as excessive cognitive load may impede guided learning process [5].

Table 1 : Principles for Reducing Extraneous Processing

| Principle | Description |
|---|---|
| Coherence | Removal of extra information which is not necessary for learning |
| Signaling | Inclusion of cues to facilitate the organization of information |
| Spatial Contiguity | Placing of related representation (i.e., words, diagrams) close to one another |

In addition to these principles, we sourced for potential enhancements to be included in the remixed simulation by reviewing the existing energy simulation [19]. Lastly, we also included features that support the inquiry-based activities. See Figure 1 and Table 2 for the features of the remixed simulation. Some of the design features outlined in Table 2 can be found in Figure 1 as indicated by the corresponding number.

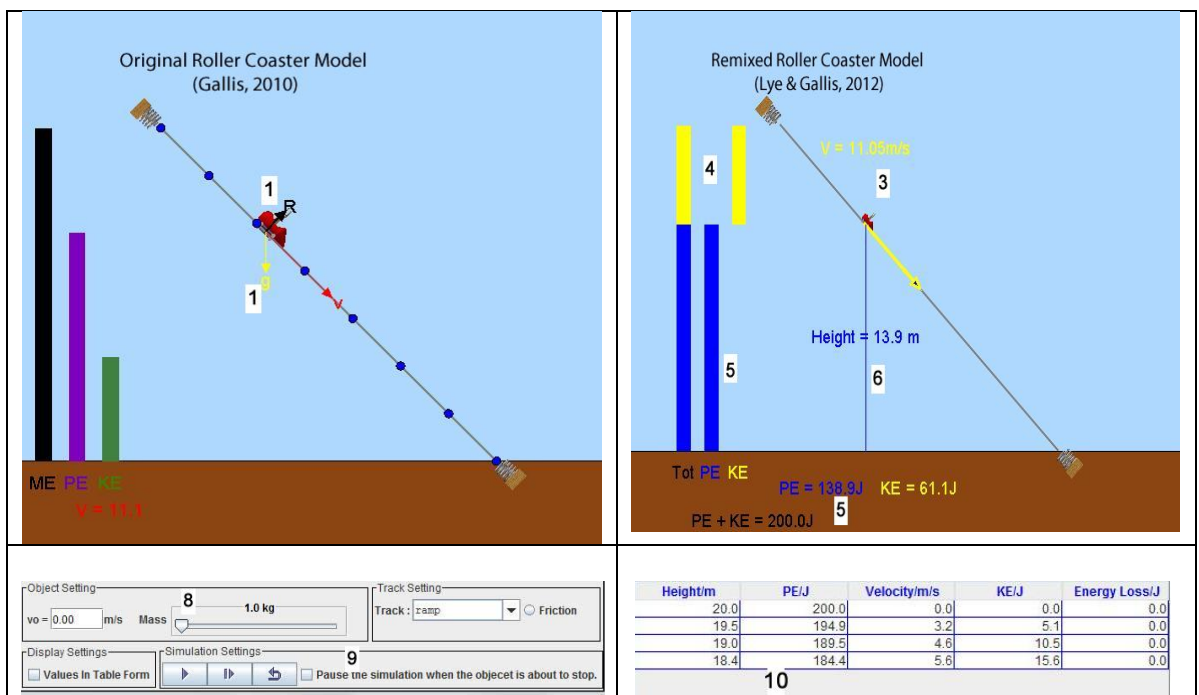

Figure 1 : Design Features of Roller Coaster

Table 2 : Design Features of the remixed simulation

| No | Design Features | Rationale |
|---|---|---|
| 1 | Removal of extra information like g (acceleration due to gravity) and R (normal force) | Coherence Principle. |
| 2 | Consistent colour scheme for concepts related to potential energy (PE) and kinetic energy (KE). | Signaling Principle. Make the relationship between the related concepts more explicit . |
| 3 | Display of velocity information in symbolic (length of arrow) and numerical form near the object. | Spatial Contiguity Effect Better visualization how the velocity changes as the object moves. |
| 4 | Modification of Total Energy bar such it is made up of KE and PE energy bars Change the orientation of KE bar such that the . | Spatial Contiguity Effect Better visualization of the total energy concept as the total energy bar is now dynamically made up of KE and PE energy bars. |
| 5 | Representation of kinetic energy, potential energy in bars and the total energy in numerical form. | Multiple representations to guide the students in inquiry-based activities like evaluating and communication of findings, responding to questions . |
| 6 | Inclusion of the height to show the relationship between height and PE | |
| 7 | Arrangement of control settings by grouping similar settings control together. | Spatial Contiguity Effect and Signaling Principle |
| 8 | Allowing the learner to pause and step through the simulation when the object is about to stop | Signaling Principle. Focus the learner's attention on what happens when the object is about to stop. |
| 9 | Allowing the learner to change the mass | Enhancement made after reviewing [19] Allows the students to design investigation to investigate how mass can affect the energy |
| 10 | Inclusion of the table of values of variables like KE | Allows the students to design experiment to investigate the relationship between these variables. |

## 4. Conclusion and Further Work

This is an ongoing case study research which is guided by the following research questions : (1) What are the inquiry learning principles (e.g., level of teacher facilitation) that can improve students' understanding of energy concepts? (2) What are the design features of the computer simulations that help the students to understand energy concepts? The first phase of the research has been completed with the energy simulation being remixed based on Mayer's principles of reducing extraneous processing, existing simulations and inquiry learning principles. The energy simulation is sent to some teachers (N=6) for reviewing. Response is largely positive with all of them agreeing that the added features are useful.

> I like the correlation between the P.E and K.E as the roller coaster gains K.E and loses P.E and vice versa on the different sets of platforms. It clearly shows the loss and gain of energy.
> The value of speed with the roller coaster (in yellow) is great because the pupils can see the motion, and see the values

In the next phase, we will work closely with the teacher to co-design the guided inquiry lesson package. Data from multiple data sources are collected. The data collection instruments include (a) teachers' feedback on the simulation (b) field notes based on classroom observations (c) student-produced artifacts (d) students' feedback on simulation , (e) students' results and (f) teacher's reflection on lesson.

**Acknowledgements**

We wish to express our deepest gratitude to the following group of people or organizations for making this research possible:
(1) Francisco Esquembre, Fu-Kwun Hwang and Wolfgang Christian for their contribution to the open source physics simulation.
(2) Michael Gallis for creating and sharing his original simulation
(3) Teachers for providing us with feedback on the simulation
(4) Ministry of Education, Educational Technology Division (ETD) and National Research Funding (NRF) for supporting and funding the project: *eduLab-003 : Java Simulation Design for Teaching and Learning* for which this research is spin off from.